\documentclass[prd,superscriptaddress,twocolumn,nopreprintnumbers,floatfix,nofootinbib]{revtex4}

\usepackage{textcomp}
\usepackage{amssymb}
\usepackage{amsmath}
\usepackage{graphicx}
\usepackage{dcolumn}
\usepackage{color,units}
\usepackage{lineno}
\usepackage{xspace}
\usepackage{mathtools}
\usepackage{physics}
\usepackage{tensor}
\usepackage{subcaption}
\usepackage[dvipsnames]{xcolor}
\usepackage[normalem]{ulem}
\usepackage[colorlinks=true,linkcolor=blue,citecolor=blue,urlcolor=blue]{hyperref}%

\newcommand{\SPA}{School of Physics and Astronomy, Monash University, Vic 3800, Australia}
\newcommand{\OzGravMonash}{OzGrav: The ARC Centre of Excellence for Gravitational Wave Discovery, Clayton VIC 3800, Australia}

\begin{document}

\title{Temperature dependent appearance of exotic matter makes nascent neutron stars spin faster}
\author{Francisco Hernandez Vivanco}
\email{francisco.hernandezvivanco@monash.edu}
\affiliation{\SPA}
\affiliation{\OzGravMonash}

\author{Paul D. Lasky}
\author{Eric Thrane}
\author{Rory Smith}
\affiliation{\SPA}
\affiliation{\OzGravMonash}

\author{Debarati Chatterjee}
\affiliation{Inter-University Centre for Astronomy and Astrophysics,
Pune University Campus, Pune 411007, India}

\author{Sarmistha Banik}
\affiliation{BITS Pilani, Hyderabad Campus, Dept of Physics, Hyderabad 500078, India }

\author{Theo Motta}
\author{Anthony Thomas}
\affiliation{CSSM and ARC Centre of Excellence for Dark Matter Particle Physics,
Department of Physics, University of Adelaide, SA 5005 Australia}

\date{\today}

\begin{abstract}
Neutron stars offer the opportunity to study the behaviour of matter at densities and temperatures inaccessible to terrestrial experiments. Gravitational-wave observations of binary neutron star coalescences can constrain the neutron-star equation of state before and after merger. After the neutron star binary merges, hyperons can form in the remnant, changing the behaviour of the neutron-star equation of state. 
In this study, we use finite-entropy equations of state to show that a post-merger remnant can spin up due to cooling. 
The magnitude of the spin-up depends on the  neutron-star equation of state.
If hyperons are present, the post-merger spin-up changes the peak gravitational-wave frequency by \unit[$\sim 540$]{Hz}, when the  entropy per baryon drops from \unit[$s=2$]{$k_B$} to \unit[$s=0$]{$k_B$}. 
If hyperons are not present, the post-merger spin-up changes by $\sim\unit[360]{Hz}$, providing a gravitational-wave signature for exotic matter. 
We expect the same qualitative behaviour whenever temperature dependent phase transitions are triggered.
\end{abstract}.

\maketitle
\section{Introduction}
The study of neutron stars using gravitational-wave observations has opened a new window to study how matter behaves at supranuclear densities. The behaviour of neutron-star matter is described by the nuclear equation of state (EoS), which determines the relation between parameters such as the mass and radius of a neutron star. The measurement of the first binary neutron star merger, GW170817, allowed us to constrain the equation of state  by measuring the tidal deformability of each neutron star~\citep{LSC_GW170817}. These results have been combined with GW190425~\citep{LSC_GW190425}, and measurements from electromagnetic observations and nuclear theory~\citep[e.g.][]{Raaijmakers_2019,Capano_2020,Dietrich_2020,hernandez_vivanco_2020}.

At sufficiently high densities, strangeness containing matter may appear in the core, e.g., in the form of hyperons or deconfined quark matter,  significantly changing the behaviour of the star. 
Signatures of strange matter can be inferred during the inspiral and post-merger of a binary neutron star system. \citet{Chatziioannou_2020} show that, during the inspiral, hadron-quark phase transitions can be detected with 50-100 observations assuming a LIGO-Virgo detector network, as long as the phase transition is strong and occurs within the detected population. 

If the appearance of strange matter does not occur before or during the inspiral, it can still be triggered after the binary merger. Since the remnant is hot after the merger, the density of the core can reach levels which cannot be reached during the inspiral. These conditions can trigger the appearance of exotic matter such as hyperons or deconfined quarks, which can be measured using gravitational-wave observations~ \citep{Sekiguchi_2011,Radice_2017,Weih_2020}. \citet{Bauswein_2019} show that hadron-quark phase transitions can be detected during the post-merger by comparing the dominant gravitational-wave frequency $f_{\mathrm{peak}}$ with the tidal deformability $\lambda$ measured during the inspiral. If no phase transition to quark matter occurs, the peak frequency depends predominantly on $\lambda$~\citep{Bauswein_2012,Bauswein_2012_b}. If $f_{\mathrm{peak}}$ shifts from the value predicted from the tidal deformability inferred from the inspiral, one may infer the existence of exotic phases.

\citet{Weih_2020} outlines four different outcomes of a post-merger remnant which depend on whether phase transitions to quark matter are triggered after the merger. These scenarios are shown in Fig.~1 of Ref.~\citep{Weih_2020} and correspond to the following outcomes: (1) a phase transition does not occur, (2) a phase transition occurs immediately after the merger, (3) a phase transition is not immediately triggered after the merger, but when it is triggered, the post-merger remnant collapses to a black hole, and (4) a phase transition is not immediately triggered, but when it is triggered, the remnant does not collapse to a black hole and forms a metastable object emitting gravitational waves at higher frequencies than it would without a phase transition.

Scenarios (1), (2) and (3) have been studied in the literature \citep[e.g.][]{Bauswein_2012,Takami_2014,Takami_2015,Kawaguchi_2018,Most_2019}. Scenario (4), referred to as ``delayed phase transition,'' was introduced in \citet{Weih_2020} and takes place a few milliseconds after the merger due to a sudden softening of the equation of state of the core, which causes the core density to overcome a critical phase-transition density. This scenario is particularly interesting because the post-merger gravitational-wave emission may be characterised by two distinct frequencies.

In this paper, we propose a fifth type of post-merger scenario that is triggered due to the cooling of a neutron star. Using finite-entropy realistic equations of state that take into account the effect of $\Lambda$-hyperons~\cite{Banik_2014}, we find that a post-merger remnant can spin up under the right conditions. The neutron star spin up, combined with the softening of the equation of state, shifts the main gravitational-wave emission frequency $f_{\mathrm{peak}}$. When $\Lambda$-hyperons are present in the core of a neutron star, we find that $f_{\mathrm{peak}}$ changes by \unit[$\sim 540$]{Hz}. If this spin-up occurs in nature, it will occur on a timescale larger than the delayed phase transition proposed in \citet{Weih_2020}.

The remainder of this paper is organized as follows. In Sec.~\ref{sec:Method}, we explain why the post-merger remnant is expected to spin up when the entropy drops after the merger. We show the conditions that have to be satisfied in order to observe the post-merger remnant spin up using gravitational-wave observations. In Sec.~\ref{sec:results} we calculate the gravitational-wave emission frequency variation when a delayed appearance of hyperons is triggered. In Sec.~\ref{sec:Discussion}, we discuss our results, and we conclude in Sec.~\ref{sec:Conclusion}.

\section{Neutron-star spin-up}\label{sec:Method}
The radius of a neutron star can significantly change as a function of temperature when a newly born neutron star cools after the merger \citep[e.g.][]{Panda_2010,stone_2019,Nunna_2020}. 
\citet{stone_2019} show that when the temperature of a neutron star increases, hyperons appear at lower densities, consequentially changing the radius of a neutron star due to a softening of the equation of state. Figure~\ref{fig:mass_radius} shows the gravitational mass $m_G$ versus radius $R$ profile of the neutron star for several equations of state used in this study. We briefly describe the equations of state here. They were also used by \citet{Nunna_2020}.
\begin{itemize}
\item \textbf{DD2: an equation of state including nucleonic matter.}  In this model \cite{Hempel_2010}, nucleonic matter is described by an ensemble of nuclei and interacting nucleons in nuclear statistical equilibrium. While uniform nuclear matter is described by a relativistic mean field model, nuclei are described using nuclear structure calculations based on nuclear Lagrangian density. The transition from the non-uniform phase (nuclei) to uniform nuclear matter is implemented using a thermodynamically consistent description with excluded volume corrections. 

\item \textbf{BHB$\Lambda\phi$: an equation of state including $\Lambda$-hyperons.} In this model ~\cite{Banik_2014}, the non-uniform nuclear matter description of the DD2 EoS \cite{Hempel_2010} is used following the standard prescription of minimization of the free energy. As the hyperon-hyperon interaction is mediated via the non-strange vector meson $\phi$, the EoS with $\Lambda$-hyperons is represented by BHB$\Lambda \phi$.
\end{itemize}

The equations of state DD2 and BHB$\Lambda\phi$ are both evaluated at entropy per baryon of \unit[$s=2$]{$k_B$} and \unit[$s=0$]{$k_B$}. From Fig.~\ref{fig:mass_radius}, we see that the radius of a \unit[2]{$M_\odot$} neutron star changes by roughly 12\% when the entropy per baryon decreases from \unit[$s=2$]{$k_B$} to \unit[$s=0$]{$k_B$}. The mass-radius relations are calculated using the numerical library \textsc{Lorene}~\citep{lorene}.

In the remainder of this section, we show that a post-merger remnant can spin up when its temperature decreases after the merger, due to the radius variation shown in Fig.~\ref{fig:mass_radius}. We calculate the increase in the main gravitational-wave emission frequency $f_\mathrm{peak}$ associated with this spin-up.

\subsection{Post-merger remnant spin up} \label{sec:spin-up}
\begin{figure}[!t]
    \centering
    \includegraphics[width=\columnwidth]{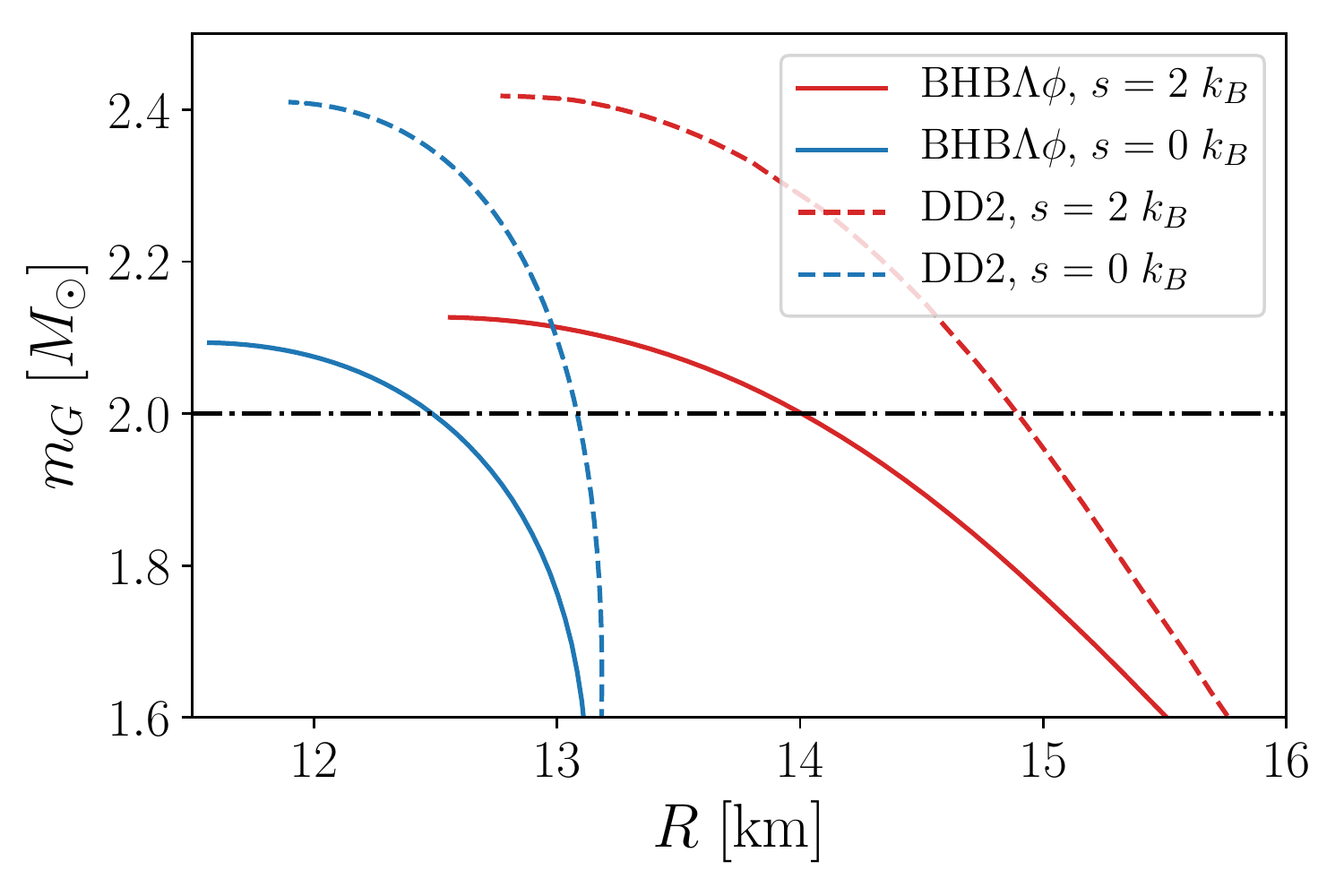}
    \caption{
    Gravitational mass versus radius relations used in this study, corresponding to the nucleonic equation of state DD2, and the BHB$\Lambda\phi$ equation of state containing $\Lambda$-hyperons. This plot shows that the radius of a \unit[2]{$M_\odot$} neutron star decreases by 12\% when the entropy drops from $s=2k_B$ to $s=0k_B$.
    }
    \label{fig:mass_radius}
\end{figure}

We first show that a post-merger remnant can spin up after the merger.  We assume that after the merger, angular momentum is conserved.
With this assumption we are assuming that the spin-up is relatively fast compared to the spin-down timescale.
We revisit this momentarily.
Thus,
\begin{equation}
    I_i \Omega_i = I_f \Omega_f.
\end{equation}
Here, $I$ is the moment of inertia, $\Omega$ is the angular velocity and the subscripts $i$ and $f$ refer to the initial and final state of the post-merger remnant. If we additionally assume that mass is conserved during the cool down and that the density of the neutron star is uniform, we find that 
\begin{equation} \label{eq:spin-up}
    \Omega_f = \left(\frac{R_i}{R_f}\right)^2 \Omega_i.
\end{equation}
Equation (\ref{eq:spin-up}) shows that if the radius of a neutron star changes by $\sim$12\%, as in Fig.~\ref{fig:mass_radius}, its angular velocity increases by $\sim$25\%. 

\subsection{Cooling timescale} \label{sec:cooling_timescale}
The post-merger remnant can spin up after the merger due to the radius changing during cooling. However, other torques related to electromagnetic and gravitational-wave emission will spin down the neutron star~\citep[e.g.][]{Paschalidis_2012}.
Here, we explain the necessary conditions under which we expect the spin up to take place. 

Let us assume that a post-merger remnant has an initial angular momentum $J$. For simplicity, we assume that gravitational-wave spin down dominates over electromagnetic torques. Under this assumption, the change of angular momentum  over time $\dot J$ is given by
\begin{equation}\label{eq:j_dot}
    \dot J = \dot \omega I + \omega \dot I.
\end{equation}
The first term,  $\dot \omega I$, is associated with spin-down due gravitational-wave emission. The second term, $\omega \dot I$, is associated with spin-up due to the radius changing with the falling temperature of the remnant after the merger. 
The post-merger remnant spin-down, caused by $\dot \omega I$ in Eq.~(\ref{eq:j_dot}), occurs during a gravitational-wave timescale $t_\mathrm{gw}$, which can be approximated by~\cite{Paschalidis_2012}

\begin{equation}\label{eq:gravitational_timescale}
    t_{\mathrm{gw}} \simeq 200 \left( \frac{\epsilon}{0.5} \right)^{-4} \left( \frac{e}{0.75} \right)^{-2} \left( \frac{R}{20 \mathrm{ km}} \right)^4 \left( \frac{m_G}{2.8 M_\odot} \right)^{-3} \mathrm{ms} .
\end{equation}
Here, $\epsilon$ is the ratio of the star's angular frequency to the break-up angular frequency, $e$ is the ellipticity of the post-merger remnant, $R$ is the radius and $m_G$ is the gravitational mass. 

The characteristic timescale for the spin-up, caused by the term $\omega \dot I$ in Eq.~(\ref{eq:j_dot}), is the cooling timescale $t_{\mathrm{cool}}$.
The dominant cooling mechanism is neutrino emission~\cite{Paschalidis_2012}:
\begin{equation}\label{eq:cooling_timescale}
    t_{\mathrm{cool}} \simeq 400 \left( \frac{m_G}{2.8M_\odot} \right) \left( \frac{R}{20 \mathrm{ km}} \right)^{-1} \left( \frac{E_\nu}{10 \mathrm{ MeV}} \right)^2 \mathrm{ms} .
\end{equation}
Here, $E_\nu$ is the root mean squared (RMS) neutrino energy. At densities of $\unit[\gtrsim  10^{11}]{g\, cm^{-3}}$ neutrinos are trapped.
Therefore the cooling timescale is predominantly determined by how long it takes for neutrinos to diffuse out of the remnant~\cite{Paschalidis_2012}. The neutrino energy $E_\nu$ (and therefore the cooling timescale) depends on the cooling transfer mechanism which is not well understood.

The condition required to observe the post-merger remnant spin up is given by
\begin{equation}\label{eq:condition}
    t_{\mathrm{cool}} < t_{\mathrm{gw}}.
\end{equation}
Equation~(\ref{eq:condition}) implies that if the remnant spin-up takes place with a timescale greater than the gravitational-wave timescale, it is suppressed by the spin-down caused by gravitational-wave emission. 
On the other hand, if $t_{\mathrm{cool}} < t_{\mathrm{gw}}$, the remnant spins up faster than it spins-down. 

\begin{figure} 
\centering
\begin{subfigure}[b]{0.5\textwidth} 
   \includegraphics[width=\columnwidth]{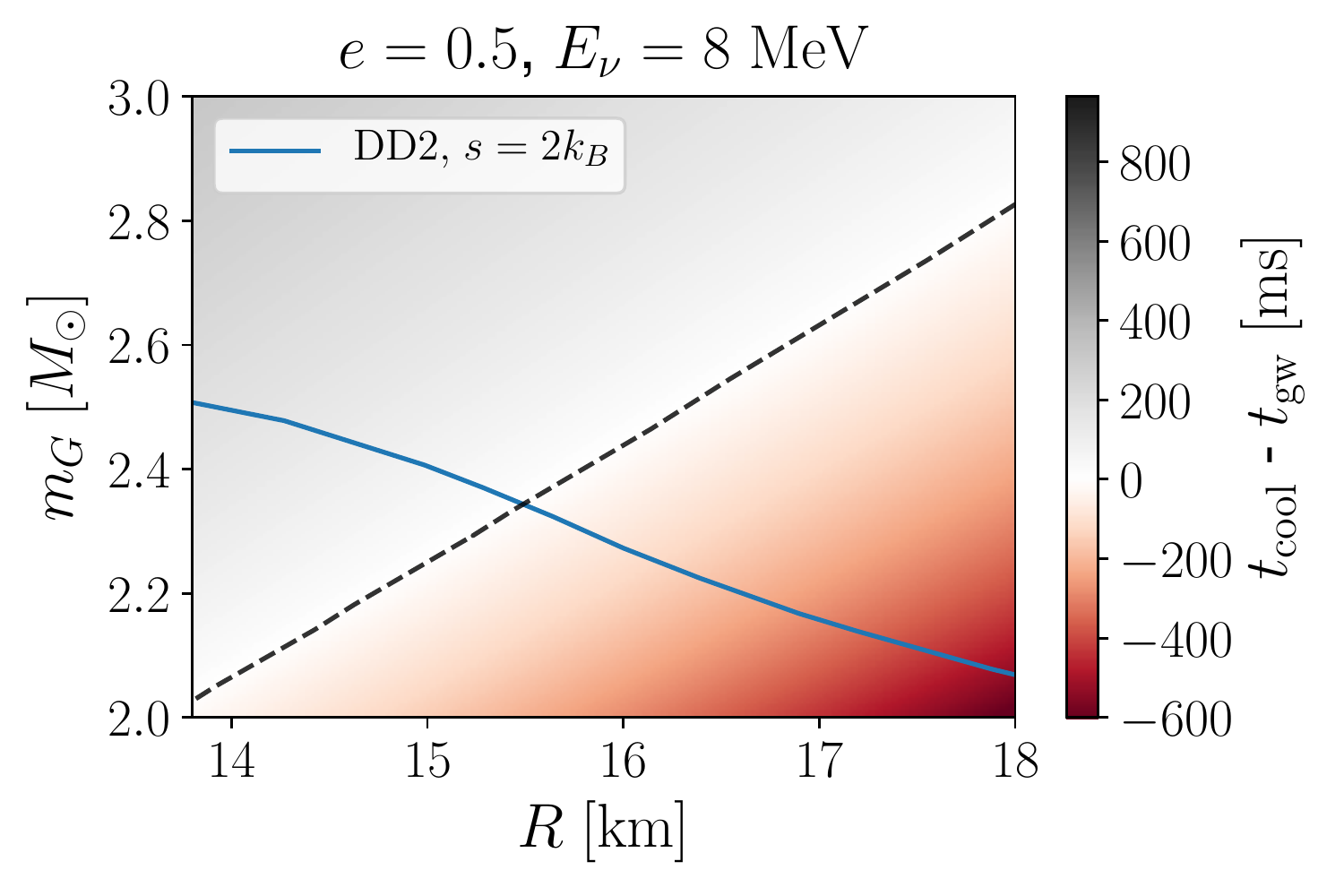}
   \caption{}
\end{subfigure}

\begin{subfigure}[b]{0.5\textwidth}
   \includegraphics[width=\columnwidth]{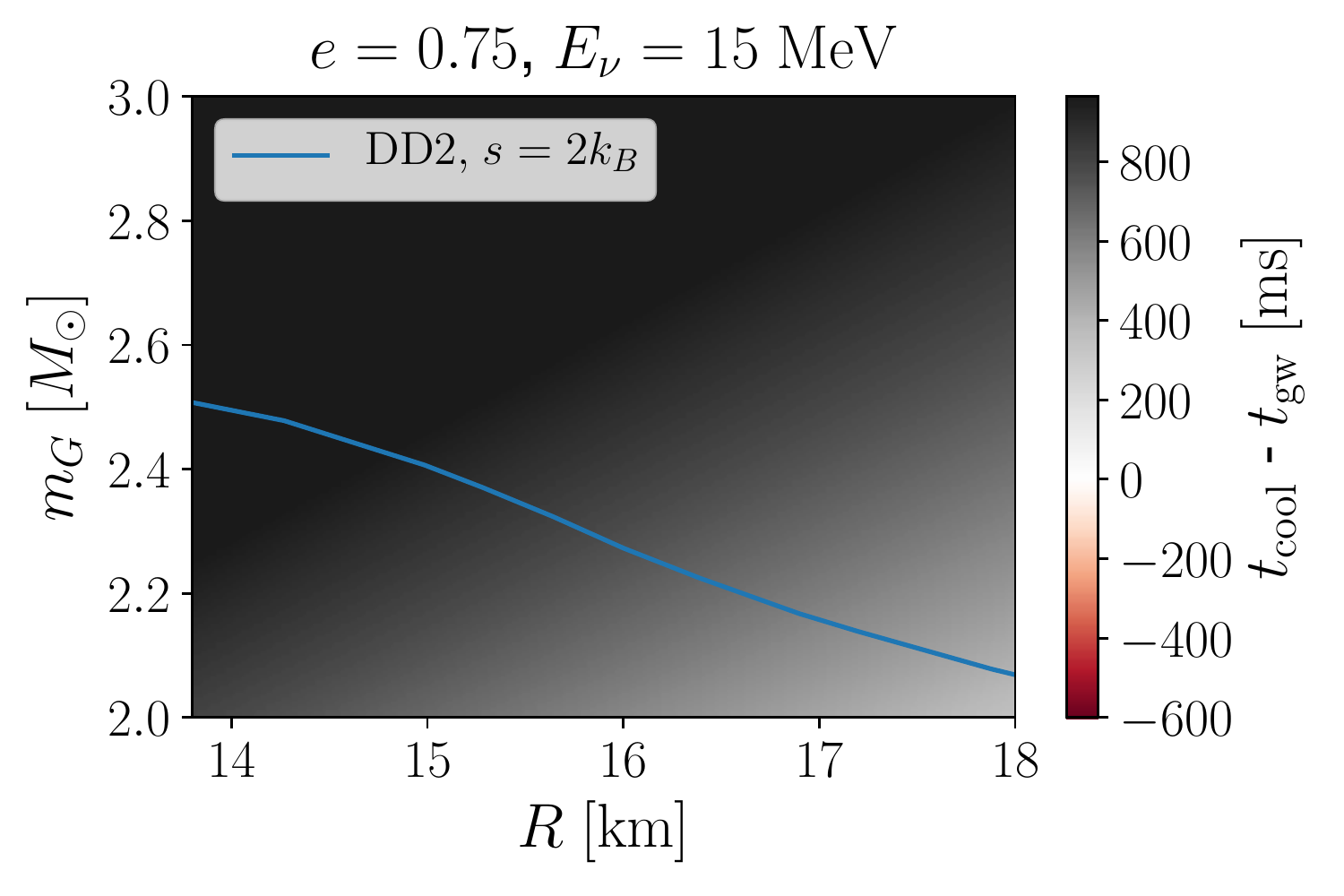}
   \caption{}
\end{subfigure}

\caption[]{(a) Spin up timescales as a function of gravitational mass and radius for a fixed ellipticity $e=0.5$ and neutrino energy \unit[$E_\nu=8$]{MeV}. Only the regions where $t_{\mathrm{cool}}- t_{\mathrm{gw}}<0$, shown in red, correspond to an observable spin up of the post-merger remnant. The blue curve corresponds to the DD2 equation of state considered in Fig.~\ref{fig:mass_radius} at \unit[$s=2$]{$k_B$} rotating with an angular momentum \unit[$J=2.2$]{$GM_\odot^2/c$}. The black dotted line corresponds to the values of masses and radii where $t_{\mathrm{gw}} = t_{\mathrm{cool}}$ (b) Same as (a) but for an ellipticity $e=0.75$ and neutrino energy \unit[$E_\nu=15$]{MeV}.}
\label{fig:fixed_e_and_E}
\end{figure}

We explore the mass-radius, and the neutrino energy-ellipticity parameter space that  satisfy $t_{\mathrm{cool}} - t_{\mathrm{gw}} < 0$ assuming $\epsilon=0.5$.
We consider two values of ellipticity and neutrino energy to determine the values of gravitational mass and radius that would result in a post-merger remnant spin up. The results are shown in Fig.~\ref{fig:fixed_e_and_E}.
We indicate in red the gravitational masses and radii that result in an observable spin-up. The black regions show the gravitational mass and radius values where the spin up is not observable. From Fig.~\ref{fig:fixed_e_and_E}, we see that for neutrino energies in the order of \unit[8]{MeV} and ellipticities in the order of $e=0.5$, the remnant spin-up is observable for masses \unit[$m_G \lesssim 2.3$]{$M_\odot$}. However, for larger neutrino energies, i.e. \unit[$E_\nu =15$]{MeV}, the remnant spin up is not be observable regardless of its mass. 

\begin{figure} 
\centering
\includegraphics[width=\columnwidth]{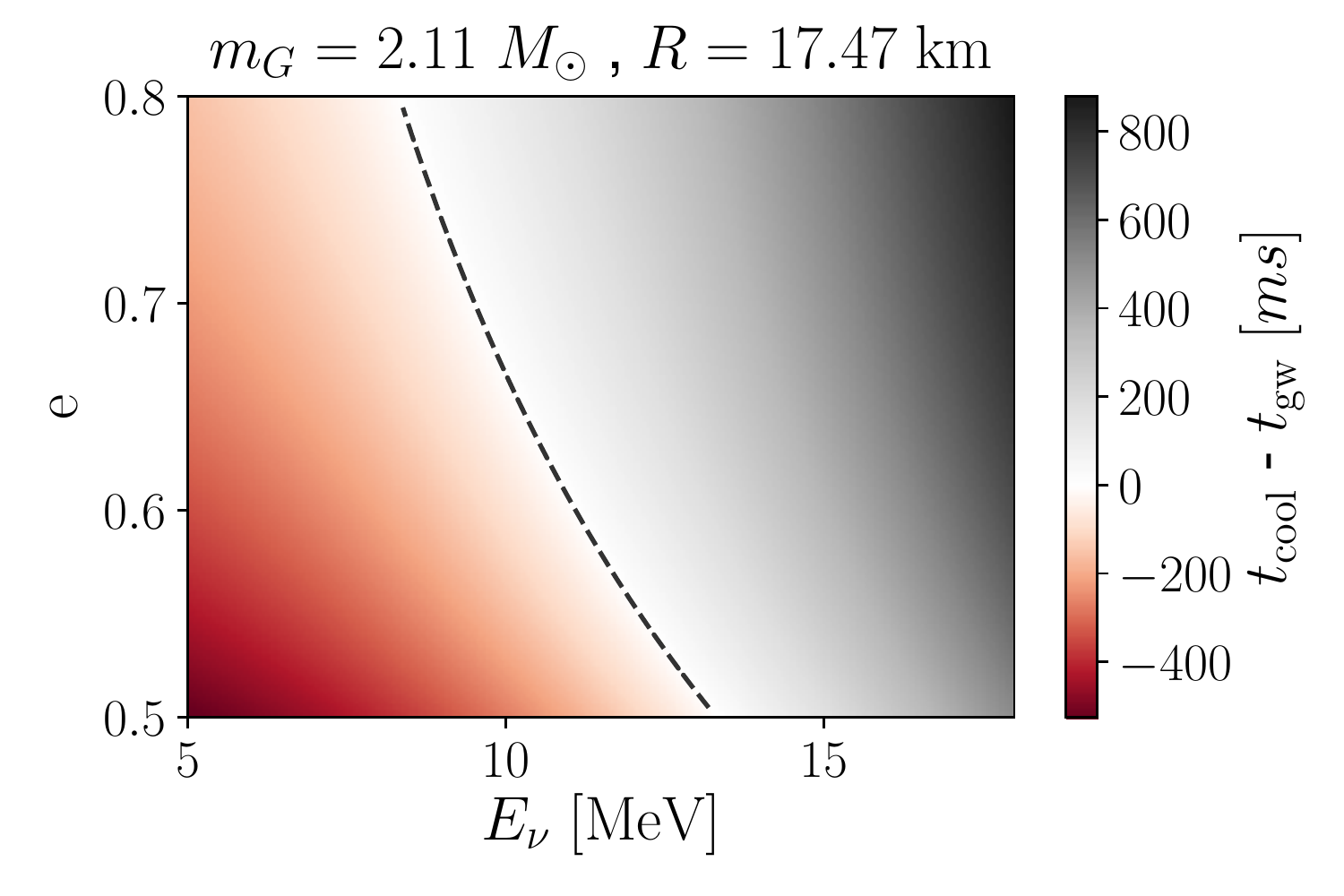}
\caption[]{Spin up timescales as a function of ellipticity and neutrino energy of a post-merger remnant of mass \unit[$m_G=2.11$]{$M_\odot$}. The remnant follows the DD2 equation of state at entropy \unit{$s=2$}{$k_B$} rotating with an angular momentum \unit[$J=2.2$]{$GM_\odot^2/c$}. Only the regions where $t_{\mathrm{cool}}- t_{\mathrm{gw}}<0$, shown in red, correspond to an observable spin up of the post-merger remnant. The black dotted curve corresponds to the values of ellipticity and neutrino energy  where $t_{\mathrm{gw}} = t_{\mathrm{cool}}$.
}
\label{fig:fixed_m_and_r}
\end{figure}

In Fig.~\ref{fig:fixed_m_and_r}, we show the values of ellipticity and neutrino energy that would be needed to observe the remnant spin up, assuming the mass of a \unit[$m_G=2.11$]{$M_\odot$} remnant assuming the DD2 equation of state at \unit[$s=2$]{$k_B$} rotating with an angular momentum \unit[$J=2.2$]{$GM_\odot^2/c$}. We see that RMS neutrino energies \unit[$\lesssim 14$]{MeV} result in an observable spin up.

\subsection{Gravitational-wave emission frequency}\label{sec:emission_frequency}
The gravitational-wave emission frequency of a post-merger remnant is proportional to the rotational frequency~\citep{Gaertig_2011,Doneva_2013}. Therefore, if the post-merger remnant spins up, the gravitational-wave emission frequency increases.
Simulations of the evolution of post-merger remnants reveal gravitational-wave spectra with characteristic peaks related to oscillation modes~\cite{Bauswein_2012,Bauswein_2012_b,Bauswein_2016,Paschalidis_2017}. The dominant frequency $f_{\mathrm{peak}}$ (also referred as $f_2$), is related to the co-rotating $l=m=2$ $f$-mode moving at a positive pattern speed in the prograde direction~\cite{Stergioulas_2011,Takami_2015,Doneva_2015}. This mode depends on the mass, radius and rotational frequency of the remnant.  Thus if we can accurately measure $f_{\mathrm{peak}}$, we can infer properties of the equation of state.

Gaertig and Kokkotas~\cite{Gaertig_2008,Gaertig_2011} describe relations for the co-rotating  $l=|m|=2$ $f$-modes in the Cowling approximation, which assumes that the spacetime remains frozen during the time evolution.  \citet{Doneva_2013} expands on the work of Gaertig and Kokkotas using realistic equations of state. The Cowling approximation does not accurately predict the $f$-mode and can include errors up to 20-30\%. Since our aim is to obtain an order of magnitude estimate of $f_\mathrm{peak}$, we approximate $f_\mathrm{peak}$  following Refs.~\cite{Gaertig_2008,Gaertig_2011,Doneva_2013}.

The $l=m=2$ $f$-mode stable branch in the co-rotating frame $\sigma_{\mathrm{corot}}$ is given by \citep{Doneva_2013}

\begin{equation}\label{eq:f_mode}
    \frac{\sigma_{\mathrm{corot}}}{\sigma_0} = 1 - 0.235\left(\frac{\Omega}{\Omega_k} \right) -0.358\left(\frac{\Omega}{\Omega_k} \right)^2,
\end{equation}
where $\Omega$ is the neutron star's rotation frequency, $\Omega_k$ is the Keplerian rotation frequency (or mass-shedding limit) and $\sigma_0$ is the $f$-mode of a non-rotating neutron star. An approximate value of the Keplerian frequency can be found in Refs.~\cite{Glendenning_1992,Haensel_1989,Lasota_1996,Friedman_1989,Doneva_2013}. We use the Keplerian approximation presented in Ref.~\cite{Doneva_2013},
\begin{equation}\label{eq:kepler_limit}
    \frac{1}{2\pi}\Omega_k \mathrm{[kHz]} =  1.716\sqrt{\frac{\bar{m}_0}{\bar{R}_0^3}} - 0.189.
\end{equation}
Here, $\bar{m}_0 = m_G / 1.4M_\odot$ and $\bar{R}_0 = R / \unit[10]{km}$ is the mass and radius in the non-rotating configuration. Equation (\ref{eq:kepler_limit}) is not a precise estimate. Its true value depends, among other parameters, on the presence of hyperons and the cooling process of the remnant. 
Additionally, the $l=2$ $f$-mode of a non-rotating neutron star $\sigma_0$ is given by~\cite{Doneva_2013}
\begin{equation}
    \frac{1}{2\pi}\sigma_0 \mathrm{[kHz]} = 1.562 +1.151  \sqrt{\frac{\bar{m}_0}{\bar{R}_0^3}}. 
\end{equation}
Finally, the co-rotating frequency $\sigma_{\mathrm{corot}}$ can be transformed to the inertial frame $\sigma_{\mathrm{inertial}}$,
\begin{equation} \label{eq:inertial_frame}
    \sigma_{\mathrm{inertial}} = \sigma_{\mathrm{corot}} - m\Omega.
\end{equation}
In the convention used in \citet{Doneva_2013}, the $m<0$ modes are prograde, i.e., the $f$-mode frequency in the inertial frame increases when the angular velocity $\Omega$ increases. We set $m=-2$ in Eq.~(\ref{eq:inertial_frame}).

\section{Results}\label{sec:results}
\begin{figure}[!t]
    \centering
    \includegraphics[width=\columnwidth]{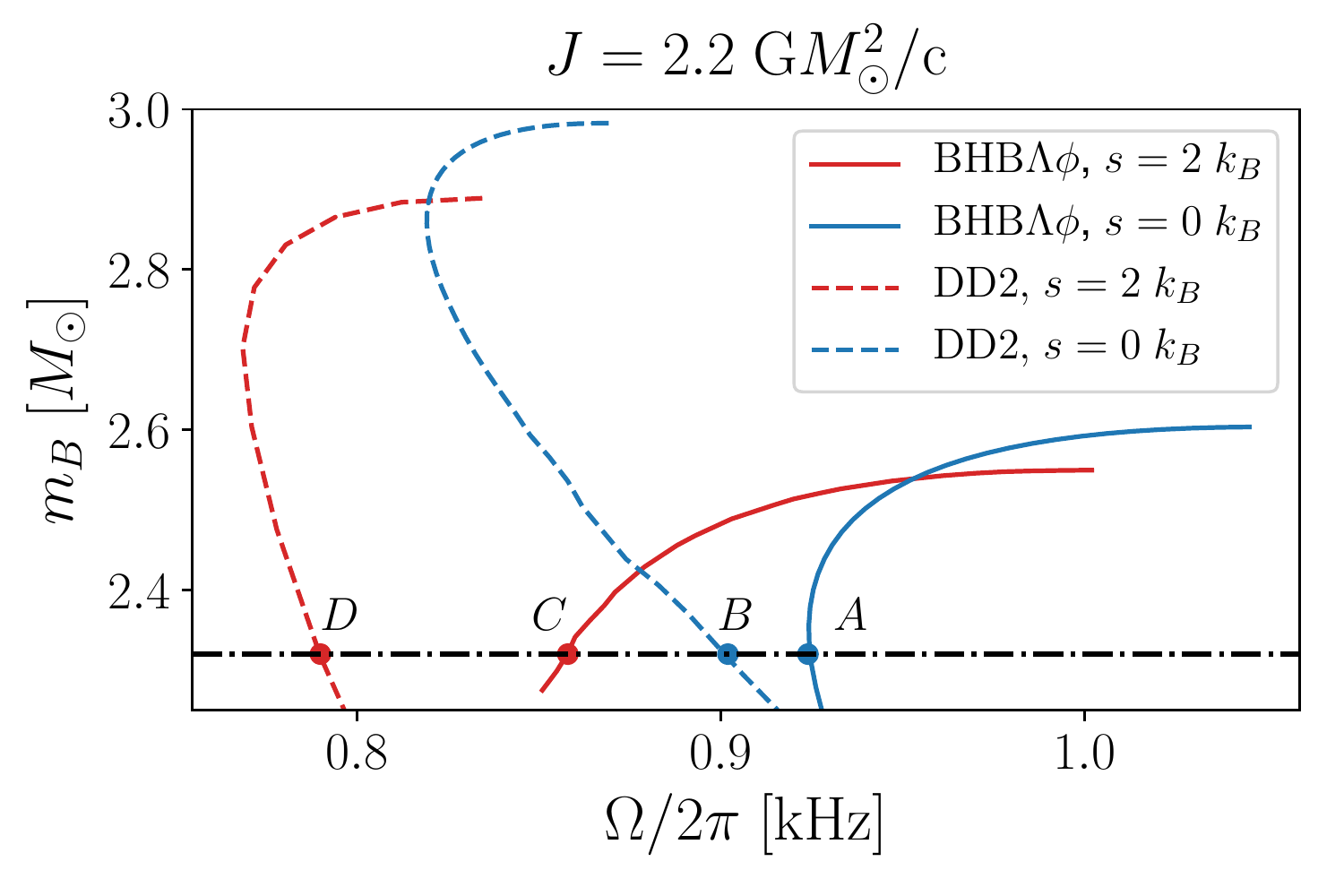}
    \caption{Baryon mass $m_B$ and rotational frequency $\Omega/2\pi$ sequences of the equations of state BHB$\Lambda\phi$ (with hyperons) and DD2 (pure nucleonic matter). We rotate the equations of state shown in Fig.~\ref{fig:mass_radius} with a constant angular momentum \unit[$J=2.2$]{G$M_\odot^2/c$}. The points $A$, $B$, $C$ and $D$ correspond to the angular rotational frequencies of an \unit[$m_B=2.32$]{$M_\odot$} post-merger remnant. If a delayed appearance of hyperons is triggered, the post-merger remnant cools down from $D \xrightarrow{} B \xrightarrow{} A $, i.e., the remnant first cools from a hot nucleonic EoS to a cold nucleonic EoS, and then hyperons appear in the core. If hyperons do not appear, the post-merger remnant cools down from $D \xrightarrow{} B$.
    }
    \label{fig:rotating_mass_radius}
\end{figure}

The post-merger remnant spin-up due to cooling described in Sec.~\ref{sec:spin-up} can happen regardless of whether  $\Lambda$-hyperons appear in the core. However, we find the peak frequency $f_\mathrm{peak}$, approximated in Sec.~\ref{sec:emission_frequency}, is emitted at a different frequency when hyperons appear during the post-merger. 
In Fig.~\ref{fig:rotating_mass_radius}, we present different scenarios that can take place after a binary neutron star merges. The points $A$, $B$, $C$ and $D$ in Fig.~\ref{fig:rotating_mass_radius} correspond to the angular velocities of an \unit[$m_B=2.32$]{$M_\odot$} post-merger remnant assuming an angular momentum \unit[$J=2.2$]{$GM_\odot^2/c$}. Here, $m_B$ is the baryon mass of a post-merger remnant which is conserved during cooling, and \unit[$m_B=2.32$]{$M_\odot$} corresponds to \unit[$m_G=2.11$]{$M_\odot$} assuming the DD2 equation of state at \unit[$s=2$]{$k_B$}.  These values are calculated using \textsc{Lorene}~\citep{lorene}.

The post-merger remnant scenarios are described as follows:

\begin{itemize}
    \item \textit{No appearance of $\Lambda$-hyperons.} The post-merger remnant cools down from $D \xrightarrow{} B $ from a hot nucleonic neutron star to a cold nucleonic one.
    
    \item \textit{Prompt appearance of $\Lambda$-hyperons.} The post-merger remnant cools down from $D \xrightarrow{} C $; hyperonic matter is present soon after the merger. 
    
    \item \textit{Delayed appearance of hyperons caused by density oscillations.} This delayed appearance of hyperons is similar to the hadron-quark phase transition presented in \citet{Weih_2020}, in which the remnant does not undergo a transition immediately after the merger.
    In this case, the post-merger remnant cools down from $D \xrightarrow{} B \xrightarrow{} A $ in Fig.~\ref{fig:rotating_mass_radius}, i.e., the remnant cools from a hot nucleonic to a cold nucleonic one, and after a few milliseconds $\Lambda$-hyperons appear. The delayed appearance of hyperons caused by density oscillation is triggered  milliseconds after the merger.
    
    \item \textit{Delayed appearance of hyperons caused by cooling.} This is the delayed appearance of hyperons proposed in this paper, which is triggered by the post-merger remnant cooling described in Sec.~\ref{sec:spin-up}. Similar to the delayed appearance of hyperons caused by density oscillations proposed in \citet{Weih_2020}, the post-merger remnant cools down from $D \xrightarrow{} B \xrightarrow{} A $, but occurs on a longer timescale that depends on how fast the post-merger remnant cools.
    The transition timescale is significantly longer than a few milliseconds. Since the delayed appearance of hyperons induced by density oscillations takes place milliseconds after the merger \citep[e.g.][]{Weih_2020}, it will be distinguishable from the delayed appearance of hyperons caused by cooling.
\end{itemize}.

\begin{figure}[!t]
    \centering
    \includegraphics[width=\columnwidth]{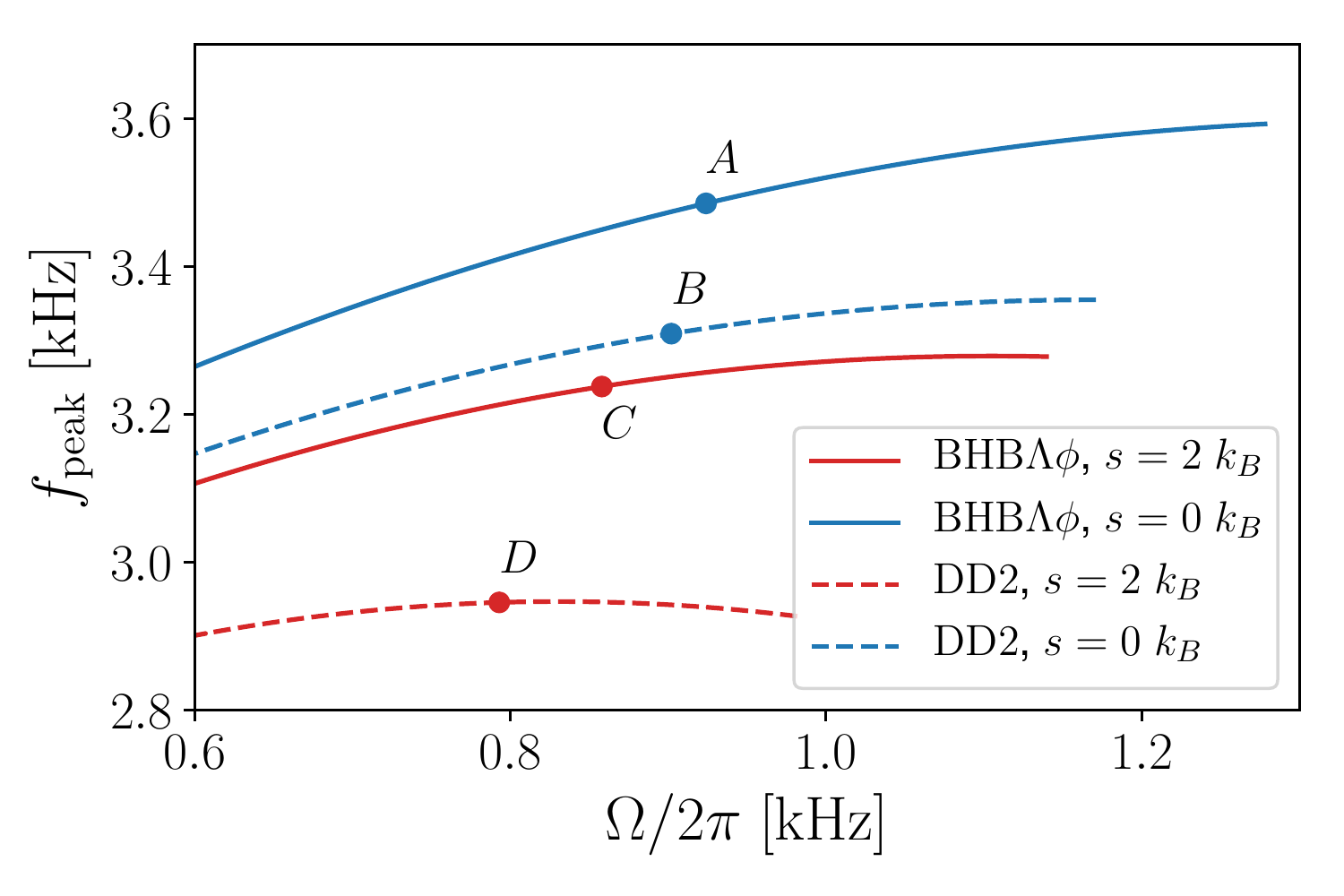}
    \caption{
    Post-merger peak frequency $f_{\mathrm{peak}}$ as a function of rotation frequency $\Omega/2\pi$ of a \unit[$m_B$=2.32]{$M_\odot$} neutron star rotating with a constant angular momentum \unit[$J=2.2$]{G$M_\odot^2/c$}. If a delayed appearance of of hyperons is triggered by cooling, the post-merger remnant spins up from $D \xrightarrow{} B \xrightarrow{} A $. In this case, $f_\mathrm{peak}$ changes by \unit[$\sim 540$]{Hz}. If hyperons do not appear, the post-merger remnant spins up from $D \xrightarrow{} B $, where $f_\mathrm{peak}$ changes by \unit[$\sim 360$]{Hz}.
    }
    \label{fig:frequency_shift}
\end{figure}

Using Eqs. (\ref{eq:f_mode}) and (\ref{eq:inertial_frame}), we calculate the $l=m=2$ $f$-mode, which corresponds to the peak frequency of a \unit[$m_B=2.32$]{$M_\odot$} post-merger remnant using the equations of state presented in \citet{Nunna_2020}.
In Fig.~\ref{fig:frequency_shift}, we plot $f_\text{peak}$ versus $\Omega$.
The points $A$, $B$, $C$ and $D$ correspond to the same points shown in Fig.~\ref{fig:rotating_mass_radius}. If hyperons do not appear after the merger, the post-merger remnant cools down from $D \xrightarrow{} B$. In this case, Fig.~\ref{fig:frequency_shift} shows that the peak frequency changes by \unit[$\sim 360$]{Hz}.  
In contrast, if a delayed appearance of hyperons is triggered, the post-merger remnant cools down from $D \xrightarrow{} B \xrightarrow{} A $.  In this case, Fig.~\ref{fig:frequency_shift} shows that the peak frequency changes by \unit[$\sim 540$]{Hz}. The peak frequency increase will allow us to probe if exotic states of matter are produced after the merger.

\section{Discussion}\label{sec:Discussion}
\citet{Chatziioannou_2017} and \citet{Easter_2020} show that gravitational-wave observations of a post-merger remnant  can resolve $f_\mathrm{peak}$ with an accuracy of \unit[20-50]{Hz} at matched-filter signal-to-noise ratio (SNR) 10, depending on the equation of state. Given that the variation of $f_\mathrm{peak}$ in a delayed appearance of hyperons is \unit[$\sim 540$]{Hz}, we expect this frequency variation to be measurable by gravitational-wave detectors.  Although measuring a post-merger signal with SNR$\sim10$ is challenging with current detectors, dedicated high-frequency detectors will help make possible such a high SNR~\citep[e.g.][]{Martynov_2019,Ackley_2020}.
For example, on average, an SNR=10 event is expected to be observed every $\sim$$\unit[4-8]{years}$ with a network of Neutron Star Extreme Matter Observatories~\citep{Ackley_2020}.

The evolution of the gravitational-wave frequency caused by cooling will be subtle, depending on precise details of cooling as well as the interplay between angular momentum loss from gravitational waves, electromagnetic radiation, and internal dissipation such as viscosity~\citep[e.g.][]{Alford_2018}.  Understanding the evolution of $f_\mathrm{peak}$ will require complex numerical simulations that take into account all of these effects as well as general relativity. 

Performing numerical simulations of the delayed appearance of hyperons, which take into account the microphysics such as magnetic fields and neutrino emission is challenging.
Numerical-relativity simulations that take into account phase transitions and appearance of $\Lambda$-hyperons are usually used to study the first \unit[$\sim 20$]{s} after the merger ~\citep[e.g.][]{Sekiguchi_2011_2,Perego_2019,Most_2019,Weih_2020}. For example, \citet{Sekiguchi_2011}  simulate transitions from nucleonic matter to hyperons using finite-entropy equations of state taking into account neutrino emission. They find that for \unit[1.35]{$M_\odot$}, equal-mass binaries, the appearance of $\Lambda$-hyperons causes the post-merger remnant to collapse to a black hole \unit[$\sim 12$]{ms} after the merger due to a sudden softening of the core. They find that the appearance of hyperons changes the gravitational-wave characteristic frequency by $\sim 20-30$\%, contrary to the nucleonic case. It remains to be seen if numerical-relativity simulations of lower mass binaries, such as masses corresponding to a total mass of \unit[$m_G \lesssim 2.3$]{$M_\odot$}, agree with the approximations used in this study.

The delayed appearance of hyperons is most readily observable if the cooling timescale $t_\mathrm{cool}$ is less than the gravitational-wave timescale $t_\mathrm{gw}$. The cooling timescale depends on the neutrino energy, which is poorly understood. In order for a post-merger remnant to spin up, we need relatively low neutrino energies, i.e., \unit[$\lesssim 14$]{MeV} assuming a \unit[$m_G=2.11$]{$M_\odot$} remnant. Refs.~\citep{Rosswog_2003,Sekiguchi_2011_2,Richers_2015,sumiyoshi2020properties} suggest that the average neutrino energy can vary between \unit[$\sim 8$]{MeV} and \unit[$\sim 30$]{MeV}. However, the models presented in Refs.~\citep{Rosswog_2003,Sekiguchi_2011_2,Richers_2015} depend sensitively on the neutrino transfer mechanism and do not account for the general theory of relativity, which can significantly alter the results.

The average neutrino energy is similarly not well constrained by observations. A search for neutrinos has been carried out on GW170817~\citep{LSC_GW170817}. Super-Kamiokande found no coincident neutrinos in the range \unit[3.5]{MeV} - \unit[100]{PeV} in two different time frames: \unit[$\pm 500$]{s} around the merger of GW170817~\citep{Abe_2018} and 14 days after the merger. Similarly, a search for high energy neutrinos in the GeV-EeV range was carried out by ANTARES, IceCube, and the Pierre Auger Observatory~\citep{Albert_2017}. They found no coincident neutrinos in the same time frames described in Ref.~\citep{Abe_2018}. However, the non-detection of neutrinos agrees with model predictions of gamma-ray bursts~\citep{Albert_2017}. 

In order to observe the delayed appearance of hyperons caused by cooling, the post-merger remnant should be long-lived and have masses \unit[$m_G \lesssim 2.3$]{$M_\odot$}.
The distribution of Galactic binary neutron star is well fit by a Gaussian with mean \unit[1.33]{$M_\odot$} and width \unit[0.09]{$M_\odot$}~\citep{Farrow_2019}. By allowing conservation of rest mass, and taking into account mass loss after the merger, the post-merger mass distribution lies in the range between \unit[2.2-2.5]{$M_\odot$}~\cite{Belczynski_2008,Lasky_2014}.\footnote{The measurement of GW190425~\citep{LSC_GW190425} shows a $\geq 5\sigma$ deviation from the Galactic distribution, although see~\citep{heavy_dns}, in which the authors use a population study to argue that GW190425 is not as different from the Milky Way population as it initially appeared. At any rate, we ignore heavier neutron stars because only low-mass binaries trigger a delayed appearance of hyperons.} In this paper, we focus on an \unit[$m_G= 2.11$]{$M_\odot$}; this limitation is given because the $f$-mode approximation defined in Eq. (\ref{eq:f_mode}) depends on the mass in the non-rotating configuration. Therefore, the maximum value we can use for our estimates is the maximum non-rotating mass (TOV mass) of the BHB$\Lambda\phi$ equation of state at $s=0$, corresponding to \unit[$m_G = 2.11$]{$M_\odot$}.

\section{Conclusion}\label{sec:Conclusion}
We use finite-entropy equations of state, which include the effect of hyperons to study the evolution of a post-merger remnant. We find that under the right circumstances, a post-merger remnant can spin up due to cooling.  The spin-up of the remnant may trigger a delayed appearance of hyperons, which can be measured by gravitational-wave observations. Since the gravitational-wave emission frequency of the post-merger is proportional to the angular velocity, we find the main gravitational-wave emission frequency  $f_{\mathrm{peak}}$ increases by \unit[$\sim 540$]{Hz} if a delayed appearance of hyperons is triggered. If hyperons do not appear and the remnant spins up, $f_\mathrm{peak}$ changes by \unit[$\sim 360$]{Hz}. This will allow us to test for exotic states of matter during the post-merger while probing protoneutron star cooling times.

The delayed appearance of hyperons is most readily observable when the cooling timescale of the post-merger is less than the gravitational-wave timescale defined by Eqs.~(\ref{eq:gravitational_timescale}) and (\ref{eq:cooling_timescale}) respectively. Assuming a \unit[$m_G=2.11$]{$M_\odot$} post-merger remnant, Eq.(\ref{eq:cooling_timescale}) is satisfied  for neutrino RMS energies \unit[$\lesssim 14$]{MeV}. 

\section*{Acknowledgments}
This work is supported through Australian Research Council Grant No. CE170100004, No. FT150100281, No. FT160100112, and No. DP180103155. F.H.V. is supported through the Monash Graduate Scholarship (MGS).

\bibliography{bibliography}

\begin{thebibliography}{51}
\expandafter\ifx\csname natexlab\endcsname\relax\def\natexlab#1{#1}\fi
\expandafter\ifx\csname bibnamefont\endcsname\relax
  \def\bibnamefont#1{#1}\fi
\expandafter\ifx\csname bibfnamefont\endcsname\relax
  \def\bibfnamefont#1{#1}\fi
\expandafter\ifx\csname citenamefont\endcsname\relax
  \def\citenamefont#1{#1}\fi
\expandafter\ifx\csname url\endcsname\relax
  \def\url#1{\texttt{#1}}\fi
\expandafter\ifx\csname urlprefix\endcsname\relax\def\urlprefix{URL }\fi
\providecommand{\bibinfo}[2]{#2}
\providecommand{\eprint}[2][]{\url{#2}}

\bibitem[{\citenamefont{Abbott et~al.}(2017)}]{LSC_GW170817}
\bibinfo{author}{\bibfnamefont{B.~P.} \bibnamefont{Abbott}}
  \bibnamefont{et~al.} (\bibinfo{collaboration}{LIGO Scientific Collaboration
  and Virgo Collaboration}), \bibinfo{journal}{Phys. Rev. Lett.}
  \textbf{\bibinfo{volume}{119}}, \bibinfo{pages}{161101}
  (\bibinfo{year}{2017}),
  \urlprefix\url{https://link.aps.org/doi/10.1103/PhysRevLett.119.161101}.

\bibitem[{\citenamefont{Abbott et~al.}(2020)}]{LSC_GW190425}
\bibinfo{author}{\bibfnamefont{B.~P.} \bibnamefont{Abbott}}
  \bibnamefont{et~al.} (\bibinfo{collaboration}{LIGO Scientific Collaboration
  and Virgo Collaboration}), \bibinfo{journal}{ApJ}
  \textbf{\bibinfo{volume}{892}}, \bibinfo{pages}{L3} (\bibinfo{year}{2020}),
  ISSN \bibinfo{issn}{2041-8213},
  \urlprefix\url{http://dx.doi.org/10.3847/2041-8213/ab75f5}.

\bibitem[{\citenamefont{Raaijmakers et~al.}(2019)\citenamefont{Raaijmakers,
  Riley, Watts, Greif, Morsink, Hebeler, Schwenk, Hinderer, Nissanke, Guillot
  et~al.}}]{Raaijmakers_2019}
\bibinfo{author}{\bibfnamefont{G.}~\bibnamefont{Raaijmakers}},
  \bibinfo{author}{\bibfnamefont{T.~E.} \bibnamefont{Riley}},
  \bibinfo{author}{\bibfnamefont{A.~L.} \bibnamefont{Watts}},
  \bibinfo{author}{\bibfnamefont{S.~K.} \bibnamefont{Greif}},
  \bibinfo{author}{\bibfnamefont{S.~M.} \bibnamefont{Morsink}},
  \bibinfo{author}{\bibfnamefont{K.}~\bibnamefont{Hebeler}},
  \bibinfo{author}{\bibfnamefont{A.}~\bibnamefont{Schwenk}},
  \bibinfo{author}{\bibfnamefont{T.}~\bibnamefont{Hinderer}},
  \bibinfo{author}{\bibfnamefont{S.}~\bibnamefont{Nissanke}},
  \bibinfo{author}{\bibfnamefont{S.}~\bibnamefont{Guillot}},
  \bibnamefont{et~al.}, \bibinfo{journal}{ApJ} \textbf{\bibinfo{volume}{887}},
  \bibinfo{pages}{L22} (\bibinfo{year}{2019}),
  \urlprefix\url{https://doi.org/10.3847%2F2041-8213%2Fab451a}.

\bibitem[{\citenamefont{Capano et~al.}(2020)\citenamefont{Capano, Tews, Brown,
  Margalit, De, Kumar, Brown, Krishnan, and Reddy}}]{Capano_2020}
\bibinfo{author}{\bibfnamefont{C.~D.} \bibnamefont{Capano}},
  \bibinfo{author}{\bibfnamefont{I.}~\bibnamefont{Tews}},
  \bibinfo{author}{\bibfnamefont{S.~M.} \bibnamefont{Brown}},
  \bibinfo{author}{\bibfnamefont{B.}~\bibnamefont{Margalit}},
  \bibinfo{author}{\bibfnamefont{S.}~\bibnamefont{De}},
  \bibinfo{author}{\bibfnamefont{S.}~\bibnamefont{Kumar}},
  \bibinfo{author}{\bibfnamefont{D.~A.} \bibnamefont{Brown}},
  \bibinfo{author}{\bibfnamefont{B.}~\bibnamefont{Krishnan}}, \bibnamefont{and}
  \bibinfo{author}{\bibfnamefont{S.}~\bibnamefont{Reddy}},
  \bibinfo{journal}{Nature Astronomy}  (\bibinfo{year}{2020}), ISSN
  \bibinfo{issn}{2397-3366},
  \urlprefix\url{https://doi.org/10.1038/s41550-020-1014-6}.

\bibitem[{\citenamefont{Dietrich et~al.}(2020)\citenamefont{Dietrich, Coughlin,
  Pang, Bulla, Heinzel, Issa, Tews, and Antier}}]{Dietrich_2020}
\bibinfo{author}{\bibfnamefont{T.}~\bibnamefont{Dietrich}},
  \bibinfo{author}{\bibfnamefont{M.~W.} \bibnamefont{Coughlin}},
  \bibinfo{author}{\bibfnamefont{P.~T.~H.} \bibnamefont{Pang}},
  \bibinfo{author}{\bibfnamefont{M.}~\bibnamefont{Bulla}},
  \bibinfo{author}{\bibfnamefont{J.}~\bibnamefont{Heinzel}},
  \bibinfo{author}{\bibfnamefont{L.}~\bibnamefont{Issa}},
  \bibinfo{author}{\bibfnamefont{I.}~\bibnamefont{Tews}}, \bibnamefont{and}
  \bibinfo{author}{\bibfnamefont{S.}~\bibnamefont{Antier}},
  \emph{\bibinfo{title}{New constraints on the supranuclear equation of state
  and the hubble constant from nuclear physics -- multi-messenger astronomy}}
  (\bibinfo{year}{2020}), \eprint{2002.11355}.

\bibitem[{\citenamefont{Hernandez Vivanco
  et~al.}(2020)\citenamefont{Hernandez Vivanco, Smith, Thrane, and
  Lasky}}]{hernandez_vivanco_2020}
\bibinfo{author}{\bibfnamefont{F.}~\bibnamefont{Hernandez Vivanco}},
  \bibinfo{author}{\bibfnamefont{R.}~\bibnamefont{Smith}},
  \bibinfo{author}{\bibfnamefont{E.}~\bibnamefont{Thrane}}, \bibnamefont{and}
  \bibinfo{author}{\bibfnamefont{P.~D.} \bibnamefont{Lasky}},
  \bibinfo{journal}{MNRAS} \textbf{\bibinfo{volume}{499}},
  \bibinfo{pages}{5972} (\bibinfo{year}{2020}), ISSN \bibinfo{issn}{0035-8711},
  \eprint{https://academic.oup.com/mnras/article-pdf/499/4/5972/34157762/staa3243.pdf},
  \urlprefix\url{https://doi.org/10.1093/mnras/staa3243}.

\bibitem[{\citenamefont{Chatziioannou and Han}(2020)}]{Chatziioannou_2020}
\bibinfo{author}{\bibfnamefont{K.}~\bibnamefont{Chatziioannou}}
  \bibnamefont{and} \bibinfo{author}{\bibfnamefont{S.}~\bibnamefont{Han}},
  \bibinfo{journal}{Phys. Rev. D} \textbf{\bibinfo{volume}{101}},
  \bibinfo{pages}{044019} (\bibinfo{year}{2020}),
  \urlprefix\url{https://link.aps.org/doi/10.1103/PhysRevD.101.044019}.

\bibitem[{\citenamefont{Sekiguchi
  et~al.}(2011{\natexlab{a}})\citenamefont{Sekiguchi, Kiuchi, Kyutoku, and
  Shibata}}]{Sekiguchi_2011}
\bibinfo{author}{\bibfnamefont{Y.}~\bibnamefont{Sekiguchi}},
  \bibinfo{author}{\bibfnamefont{K.}~\bibnamefont{Kiuchi}},
  \bibinfo{author}{\bibfnamefont{K.}~\bibnamefont{Kyutoku}}, \bibnamefont{and}
  \bibinfo{author}{\bibfnamefont{M.}~\bibnamefont{Shibata}},
  \bibinfo{journal}{Phys. Rev. Lett.} \textbf{\bibinfo{volume}{107}}
  (\bibinfo{year}{2011}{\natexlab{a}}), ISSN \bibinfo{issn}{1079-7114},
  \urlprefix\url{http://dx.doi.org/10.1103/PhysRevLett.107.211101}.

\bibitem[{\citenamefont{Radice et~al.}(2017)\citenamefont{Radice, Bernuzzi,
  Pozzo, Roberts, and Ott}}]{Radice_2017}
\bibinfo{author}{\bibfnamefont{D.}~\bibnamefont{Radice}},
  \bibinfo{author}{\bibfnamefont{S.}~\bibnamefont{Bernuzzi}},
  \bibinfo{author}{\bibfnamefont{W.~D.} \bibnamefont{Pozzo}},
  \bibinfo{author}{\bibfnamefont{L.~F.} \bibnamefont{Roberts}},
  \bibnamefont{and} \bibinfo{author}{\bibfnamefont{C.~D.} \bibnamefont{Ott}},
  \bibinfo{journal}{ApJ} \textbf{\bibinfo{volume}{842}}, \bibinfo{pages}{L10}
  (\bibinfo{year}{2017}),
  \urlprefix\url{https://doi.org/10.3847%2F2041-8213%2Faa775f}.

\bibitem[{\citenamefont{Weih et~al.}(2020)\citenamefont{Weih, Hanauske, and
  Rezzolla}}]{Weih_2020}
\bibinfo{author}{\bibfnamefont{L.~R.} \bibnamefont{Weih}},
  \bibinfo{author}{\bibfnamefont{M.}~\bibnamefont{Hanauske}}, \bibnamefont{and}
  \bibinfo{author}{\bibfnamefont{L.}~\bibnamefont{Rezzolla}},
  \bibinfo{journal}{Phys. Rev. Lett.} \textbf{\bibinfo{volume}{124}}
  (\bibinfo{year}{2020}), ISSN \bibinfo{issn}{1079-7114},
  \urlprefix\url{http://dx.doi.org/10.1103/PhysRevLett.124.171103}.

\bibitem[{\citenamefont{Bauswein et~al.}(2019)\citenamefont{Bauswein, Bastian,
  Blaschke, Chatziioannou, Clark, Fischer, and Oertel}}]{Bauswein_2019}
\bibinfo{author}{\bibfnamefont{A.}~\bibnamefont{Bauswein}},
  \bibinfo{author}{\bibfnamefont{N.-U.~F.} \bibnamefont{Bastian}},
  \bibinfo{author}{\bibfnamefont{D.~B.} \bibnamefont{Blaschke}},
  \bibinfo{author}{\bibfnamefont{K.}~\bibnamefont{Chatziioannou}},
  \bibinfo{author}{\bibfnamefont{J.~A.} \bibnamefont{Clark}},
  \bibinfo{author}{\bibfnamefont{T.}~\bibnamefont{Fischer}}, \bibnamefont{and}
  \bibinfo{author}{\bibfnamefont{M.}~\bibnamefont{Oertel}},
  \bibinfo{journal}{Phys. Rev. Lett.} \textbf{\bibinfo{volume}{122}},
  \bibinfo{pages}{061102} (\bibinfo{year}{2019}),
  \urlprefix\url{https://link.aps.org/doi/10.1103/PhysRevLett.122.061102}.

\bibitem[{\citenamefont{Bauswein and Janka}(2012)}]{Bauswein_2012}
\bibinfo{author}{\bibfnamefont{A.}~\bibnamefont{Bauswein}} \bibnamefont{and}
  \bibinfo{author}{\bibfnamefont{H.-T.} \bibnamefont{Janka}},
  \bibinfo{journal}{Phys. Rev. Lett.} \textbf{\bibinfo{volume}{108}}
  (\bibinfo{year}{2012}), ISSN \bibinfo{issn}{1079-7114},
  \urlprefix\url{http://dx.doi.org/10.1103/PhysRevLett.108.011101}.

\bibitem[{\citenamefont{Bauswein et~al.}(2012)\citenamefont{Bauswein, Janka,
  Hebeler, and Schwenk}}]{Bauswein_2012_b}
\bibinfo{author}{\bibfnamefont{A.}~\bibnamefont{Bauswein}},
  \bibinfo{author}{\bibfnamefont{H.-T.} \bibnamefont{Janka}},
  \bibinfo{author}{\bibfnamefont{K.}~\bibnamefont{Hebeler}}, \bibnamefont{and}
  \bibinfo{author}{\bibfnamefont{A.}~\bibnamefont{Schwenk}},
  \bibinfo{journal}{Phys. Rev. D} \textbf{\bibinfo{volume}{86}},
  \bibinfo{pages}{063001} (\bibinfo{year}{2012}),
  \urlprefix\url{https://link.aps.org/doi/10.1103/PhysRevD.86.063001}.

\bibitem[{\citenamefont{Takami et~al.}(2014)\citenamefont{Takami, Rezzolla, and
  Baiotti}}]{Takami_2014}
\bibinfo{author}{\bibfnamefont{K.}~\bibnamefont{Takami}},
  \bibinfo{author}{\bibfnamefont{L.}~\bibnamefont{Rezzolla}}, \bibnamefont{and}
  \bibinfo{author}{\bibfnamefont{L.}~\bibnamefont{Baiotti}},
  \bibinfo{journal}{Phys. Rev. Lett.} \textbf{\bibinfo{volume}{113}}
  (\bibinfo{year}{2014}), ISSN \bibinfo{issn}{1079-7114},
  \urlprefix\url{http://dx.doi.org/10.1103/PhysRevLett.113.091104}.

\bibitem[{\citenamefont{Takami et~al.}(2015)\citenamefont{Takami, Rezzolla, and
  Baiotti}}]{Takami_2015}
\bibinfo{author}{\bibfnamefont{K.}~\bibnamefont{Takami}},
  \bibinfo{author}{\bibfnamefont{L.}~\bibnamefont{Rezzolla}}, \bibnamefont{and}
  \bibinfo{author}{\bibfnamefont{L.}~\bibnamefont{Baiotti}},
  \bibinfo{journal}{Phys. Rev. D} \textbf{\bibinfo{volume}{91}},
  \bibinfo{pages}{064001} (\bibinfo{year}{2015}),
  \urlprefix\url{https://link.aps.org/doi/10.1103/PhysRevD.91.064001}.

\bibitem[{\citenamefont{Kawaguchi et~al.}(2018)\citenamefont{Kawaguchi, Kiuchi,
  Kyutoku, Sekiguchi, Shibata, and Taniguchi}}]{Kawaguchi_2018}
\bibinfo{author}{\bibfnamefont{K.}~\bibnamefont{Kawaguchi}},
  \bibinfo{author}{\bibfnamefont{K.}~\bibnamefont{Kiuchi}},
  \bibinfo{author}{\bibfnamefont{K.}~\bibnamefont{Kyutoku}},
  \bibinfo{author}{\bibfnamefont{Y.}~\bibnamefont{Sekiguchi}},
  \bibinfo{author}{\bibfnamefont{M.}~\bibnamefont{Shibata}}, \bibnamefont{and}
  \bibinfo{author}{\bibfnamefont{K.}~\bibnamefont{Taniguchi}},
  \bibinfo{journal}{Phys. Rev. D} \textbf{\bibinfo{volume}{97}}
  (\bibinfo{year}{2018}), ISSN \bibinfo{issn}{2470-0029},
  \urlprefix\url{http://dx.doi.org/10.1103/PhysRevD.97.044044}.

\bibitem[{\citenamefont{Most et~al.}(2019)\citenamefont{Most, Papenfort,
  Dexheimer, Hanauske, Schramm, St\"ocker, and Rezzolla}}]{Most_2019}
\bibinfo{author}{\bibfnamefont{E.~R.} \bibnamefont{Most}},
  \bibinfo{author}{\bibfnamefont{L.~J.} \bibnamefont{Papenfort}},
  \bibinfo{author}{\bibfnamefont{V.}~\bibnamefont{Dexheimer}},
  \bibinfo{author}{\bibfnamefont{M.}~\bibnamefont{Hanauske}},
  \bibinfo{author}{\bibfnamefont{S.}~\bibnamefont{Schramm}},
  \bibinfo{author}{\bibfnamefont{H.}~\bibnamefont{St\"ocker}},
  \bibnamefont{and} \bibinfo{author}{\bibfnamefont{L.}~\bibnamefont{Rezzolla}},
  \bibinfo{journal}{Phys. Rev. Lett.} \textbf{\bibinfo{volume}{122}},
  \bibinfo{pages}{061101} (\bibinfo{year}{2019}),
  \urlprefix\url{https://link.aps.org/doi/10.1103/PhysRevLett.122.061101}.

\bibitem[{\citenamefont{Banik et~al.}(2014)\citenamefont{Banik, Hempel, and
  Bandyopadhyay}}]{Banik_2014}
\bibinfo{author}{\bibfnamefont{S.}~\bibnamefont{Banik}},
  \bibinfo{author}{\bibfnamefont{M.}~\bibnamefont{Hempel}}, \bibnamefont{and}
  \bibinfo{author}{\bibfnamefont{D.}~\bibnamefont{Bandyopadhyay}},
  \bibinfo{journal}{ApJS} \textbf{\bibinfo{volume}{214}}, \bibinfo{pages}{22}
  (\bibinfo{year}{2014}),
  \urlprefix\url{https://doi.org/10.1088%2F0067-0049%2F214%2F2%2F22}.

\bibitem[{\citenamefont{Panda et~al.}(2010)\citenamefont{Panda, Provid\^encia,
  and Menezes}}]{Panda_2010}
\bibinfo{author}{\bibfnamefont{P.~K.} \bibnamefont{Panda}},
  \bibinfo{author}{\bibfnamefont{C.}~\bibnamefont{Provid\^encia}},
  \bibnamefont{and} \bibinfo{author}{\bibfnamefont{D.~P.}
  \bibnamefont{Menezes}}, \bibinfo{journal}{Phys. Rev. C}
  \textbf{\bibinfo{volume}{82}}, \bibinfo{pages}{045801}
  (\bibinfo{year}{2010}),
  \urlprefix\url{https://link.aps.org/doi/10.1103/PhysRevC.82.045801}.

\bibitem[{\citenamefont{Stone et~al.}(2019)\citenamefont{Stone, Dexheimer,
  Guichon, and Thomas}}]{stone_2019}
\bibinfo{author}{\bibfnamefont{J.~R.} \bibnamefont{Stone}},
  \bibinfo{author}{\bibfnamefont{V.}~\bibnamefont{Dexheimer}},
  \bibinfo{author}{\bibfnamefont{P.~A.~M.} \bibnamefont{Guichon}},
  \bibnamefont{and} \bibinfo{author}{\bibfnamefont{A.~W.}
  \bibnamefont{Thomas}}, \emph{\bibinfo{title}{Hot dense matter in the
  quark-meson-coupling model (qmc): Equation of state and composition of
  proto-neutron stars}} (\bibinfo{year}{2019}), \eprint{1906.11100}.

\bibitem[{\citenamefont{Nunna et~al.}(2020)\citenamefont{Nunna, Banik, and
  Chatterjee}}]{Nunna_2020}
\bibinfo{author}{\bibfnamefont{K.~P.} \bibnamefont{Nunna}},
  \bibinfo{author}{\bibfnamefont{S.}~\bibnamefont{Banik}}, \bibnamefont{and}
  \bibinfo{author}{\bibfnamefont{D.}~\bibnamefont{Chatterjee}},
  \bibinfo{journal}{ApJ} \textbf{\bibinfo{volume}{896}}, \bibinfo{pages}{109}
  (\bibinfo{year}{2020}), ISSN \bibinfo{issn}{1538-4357},
  \urlprefix\url{http://dx.doi.org/10.3847/1538-4357/ab8f2c}.

\bibitem[{\citenamefont{Hempel and Schaffner-Bielich}(2010)}]{Hempel_2010}
\bibinfo{author}{\bibfnamefont{M.}~\bibnamefont{Hempel}} \bibnamefont{and}
  \bibinfo{author}{\bibfnamefont{J.}~\bibnamefont{Schaffner-Bielich}},
  \bibinfo{journal}{Nuclear Physics A} \textbf{\bibinfo{volume}{837}},
  \bibinfo{pages}{210–254} (\bibinfo{year}{2010}), ISSN
  \bibinfo{issn}{0375-9474},
  \urlprefix\url{http://dx.doi.org/10.1016/j.nuclphysa.2010.02.010}.

\bibitem[{lor()}]{lorene}
\emph{\bibinfo{title}{\textsc{LORENE}}},
  \bibinfo{howpublished}{\url{https://lorene.obspm.fr/}}.

\bibitem[{\citenamefont{Paschalidis et~al.}(2012)\citenamefont{Paschalidis,
  Etienne, and Shapiro}}]{Paschalidis_2012}
\bibinfo{author}{\bibfnamefont{V.}~\bibnamefont{Paschalidis}},
  \bibinfo{author}{\bibfnamefont{Z.~B.} \bibnamefont{Etienne}},
  \bibnamefont{and} \bibinfo{author}{\bibfnamefont{S.~L.}
  \bibnamefont{Shapiro}}, \bibinfo{journal}{Phys. Rev. D}
  \textbf{\bibinfo{volume}{86}} (\bibinfo{year}{2012}), ISSN
  \bibinfo{issn}{1550-2368},
  \urlprefix\url{http://dx.doi.org/10.1103/PhysRevD.86.064032}.

\bibitem[{\citenamefont{Gaertig and Kokkotas}(2011)}]{Gaertig_2011}
\bibinfo{author}{\bibfnamefont{E.}~\bibnamefont{Gaertig}} \bibnamefont{and}
  \bibinfo{author}{\bibfnamefont{K.~D.} \bibnamefont{Kokkotas}},
  \bibinfo{journal}{Phys. Rev. D} \textbf{\bibinfo{volume}{83}},
  \bibinfo{pages}{064031} (\bibinfo{year}{2011}),
  \urlprefix\url{https://link.aps.org/doi/10.1103/PhysRevD.83.064031}.

\bibitem[{\citenamefont{Doneva et~al.}(2013)\citenamefont{Doneva, Gaertig,
  Kokkotas, and Kr\"uger}}]{Doneva_2013}
\bibinfo{author}{\bibfnamefont{D.~D.} \bibnamefont{Doneva}},
  \bibinfo{author}{\bibfnamefont{E.}~\bibnamefont{Gaertig}},
  \bibinfo{author}{\bibfnamefont{K.~D.} \bibnamefont{Kokkotas}},
  \bibnamefont{and} \bibinfo{author}{\bibfnamefont{C.}~\bibnamefont{Kr\"uger}},
  \bibinfo{journal}{Phys. Rev. D} \textbf{\bibinfo{volume}{88}},
  \bibinfo{pages}{044052} (\bibinfo{year}{2013}),
  \urlprefix\url{https://link.aps.org/doi/10.1103/PhysRevD.88.044052}.

\bibitem[{\citenamefont{Bauswein et~al.}(2016)\citenamefont{Bauswein,
  Stergioulas, and Janka}}]{Bauswein_2016}
\bibinfo{author}{\bibfnamefont{A.}~\bibnamefont{Bauswein}},
  \bibinfo{author}{\bibfnamefont{N.}~\bibnamefont{Stergioulas}},
  \bibnamefont{and} \bibinfo{author}{\bibfnamefont{H.-T.} \bibnamefont{Janka}},
  \bibinfo{journal}{The European Physical Journal A}
  \textbf{\bibinfo{volume}{52}}, \bibinfo{pages}{56} (\bibinfo{year}{2016}),
  ISSN \bibinfo{issn}{1434-601X},
  \urlprefix\url{https://doi.org/10.1140/epja/i2016-16056-7}.

\bibitem[{\citenamefont{Paschalidis and Stergioulas}(2017)}]{Paschalidis_2017}
\bibinfo{author}{\bibfnamefont{V.}~\bibnamefont{Paschalidis}} \bibnamefont{and}
  \bibinfo{author}{\bibfnamefont{N.}~\bibnamefont{Stergioulas}},
  \bibinfo{journal}{Living Reviews in Relativity}
  \textbf{\bibinfo{volume}{20}}, \bibinfo{pages}{7} (\bibinfo{year}{2017}),
  ISSN \bibinfo{issn}{1433-8351},
  \urlprefix\url{https://doi.org/10.1007/s41114-017-0008-x}.

\bibitem[{\citenamefont{Stergioulas et~al.}(2011)\citenamefont{Stergioulas,
  Bauswein, Zagkouris, and Janka}}]{Stergioulas_2011}
\bibinfo{author}{\bibfnamefont{N.}~\bibnamefont{Stergioulas}},
  \bibinfo{author}{\bibfnamefont{A.}~\bibnamefont{Bauswein}},
  \bibinfo{author}{\bibfnamefont{K.}~\bibnamefont{Zagkouris}},
  \bibnamefont{and} \bibinfo{author}{\bibfnamefont{H.-T.} \bibnamefont{Janka}},
  \bibinfo{journal}{MNRAS} \textbf{\bibinfo{volume}{418}},
  \bibinfo{pages}{427–436} (\bibinfo{year}{2011}), ISSN
  \bibinfo{issn}{0035-8711},
  \urlprefix\url{http://dx.doi.org/10.1111/j.1365-2966.2011.19493.x}.

\bibitem[{\citenamefont{Doneva et~al.}(2015)\citenamefont{Doneva, Kokkotas, and
  Pnigouras}}]{Doneva_2015}
\bibinfo{author}{\bibfnamefont{D.~D.} \bibnamefont{Doneva}},
  \bibinfo{author}{\bibfnamefont{K.~D.} \bibnamefont{Kokkotas}},
  \bibnamefont{and}
  \bibinfo{author}{\bibfnamefont{P.}~\bibnamefont{Pnigouras}},
  \bibinfo{journal}{Phys. Rev. D} \textbf{\bibinfo{volume}{92}}
  (\bibinfo{year}{2015}), ISSN \bibinfo{issn}{1550-2368},
  \urlprefix\url{http://dx.doi.org/10.1103/PhysRevD.92.104040}.

\bibitem[{\citenamefont{Gaertig and Kokkotas}(2008)}]{Gaertig_2008}
\bibinfo{author}{\bibfnamefont{E.}~\bibnamefont{Gaertig}} \bibnamefont{and}
  \bibinfo{author}{\bibfnamefont{K.~D.} \bibnamefont{Kokkotas}},
  \bibinfo{journal}{Phys. Rev. D} \textbf{\bibinfo{volume}{78}},
  \bibinfo{pages}{064063} (\bibinfo{year}{2008}),
  \urlprefix\url{https://link.aps.org/doi/10.1103/PhysRevD.78.064063}.

\bibitem[{\citenamefont{Glendenning}(1992)}]{Glendenning_1992}
\bibinfo{author}{\bibfnamefont{N.~K.} \bibnamefont{Glendenning}},
  \bibinfo{journal}{Phys. Rev. D} \textbf{\bibinfo{volume}{46}},
  \bibinfo{pages}{4161} (\bibinfo{year}{1992}),
  \urlprefix\url{https://link.aps.org/doi/10.1103/PhysRevD.46.4161}.

\bibitem[{\citenamefont{Haensel and Zdunik}(1989)}]{Haensel_1989}
\bibinfo{author}{\bibfnamefont{P.}~\bibnamefont{Haensel}} \bibnamefont{and}
  \bibinfo{author}{\bibfnamefont{J.~L.} \bibnamefont{Zdunik}},
  \bibinfo{journal}{Nature} \textbf{\bibinfo{volume}{340}},
  \bibinfo{pages}{617} (\bibinfo{year}{1989}), ISSN \bibinfo{issn}{1476-4687},
  \urlprefix\url{https://doi.org/10.1038/340617a0}.

\bibitem[{\citenamefont{Lasota et~al.}(1996)\citenamefont{Lasota, Haensel, and
  Abramowicz}}]{Lasota_1996}
\bibinfo{author}{\bibfnamefont{J.-P.} \bibnamefont{Lasota}},
  \bibinfo{author}{\bibfnamefont{P.}~\bibnamefont{Haensel}}, \bibnamefont{and}
  \bibinfo{author}{\bibfnamefont{M.~A.} \bibnamefont{Abramowicz}},
  \bibinfo{journal}{ApJ} \textbf{\bibinfo{volume}{456}}, \bibinfo{pages}{300}
  (\bibinfo{year}{1996}), ISSN \bibinfo{issn}{1538-4357},
  \urlprefix\url{http://dx.doi.org/10.1086/176650}.

\bibitem[{\citenamefont{Friedman et~al.}(1989)\citenamefont{Friedman, Ipser,
  and Parker}}]{Friedman_1989}
\bibinfo{author}{\bibfnamefont{J.~L.} \bibnamefont{Friedman}},
  \bibinfo{author}{\bibfnamefont{J.~R.} \bibnamefont{Ipser}}, \bibnamefont{and}
  \bibinfo{author}{\bibfnamefont{L.}~\bibnamefont{Parker}},
  \bibinfo{journal}{Phys. Rev. Lett.} \textbf{\bibinfo{volume}{62}},
  \bibinfo{pages}{3015} (\bibinfo{year}{1989}),
  \urlprefix\url{https://link.aps.org/doi/10.1103/PhysRevLett.62.3015}.

\bibitem[{\citenamefont{Chatziioannou et~al.}(2017)\citenamefont{Chatziioannou,
  Clark, Bauswein, Millhouse, Littenberg, and Cornish}}]{Chatziioannou_2017}
\bibinfo{author}{\bibfnamefont{K.}~\bibnamefont{Chatziioannou}},
  \bibinfo{author}{\bibfnamefont{J.~A.} \bibnamefont{Clark}},
  \bibinfo{author}{\bibfnamefont{A.}~\bibnamefont{Bauswein}},
  \bibinfo{author}{\bibfnamefont{M.}~\bibnamefont{Millhouse}},
  \bibinfo{author}{\bibfnamefont{T.~B.} \bibnamefont{Littenberg}},
  \bibnamefont{and} \bibinfo{author}{\bibfnamefont{N.}~\bibnamefont{Cornish}},
  \bibinfo{journal}{Phys. Rev. D} \textbf{\bibinfo{volume}{96}}
  (\bibinfo{year}{2017}), ISSN \bibinfo{issn}{2470-0029},
  \urlprefix\url{http://dx.doi.org/10.1103/PhysRevD.96.124035}.

\bibitem[{\citenamefont{Easter et~al.}(2020)\citenamefont{Easter, Ghonge,
  Lasky, Casey, Clark, Hernandez~Vivanco, and Chatziioannou}}]{Easter_2020}
\bibinfo{author}{\bibfnamefont{P.~J.} \bibnamefont{Easter}},
  \bibinfo{author}{\bibfnamefont{S.}~\bibnamefont{Ghonge}},
  \bibinfo{author}{\bibfnamefont{P.~D.} \bibnamefont{Lasky}},
  \bibinfo{author}{\bibfnamefont{A.~R.} \bibnamefont{Casey}},
  \bibinfo{author}{\bibfnamefont{J.~A.} \bibnamefont{Clark}},
  \bibinfo{author}{\bibfnamefont{F.}~\bibnamefont{Hernandez~Vivanco}},
  \bibnamefont{and}
  \bibinfo{author}{\bibfnamefont{K.}~\bibnamefont{Chatziioannou}},
  \bibinfo{journal}{Phys. Rev. D} \textbf{\bibinfo{volume}{102}}
  (\bibinfo{year}{2020}), ISSN \bibinfo{issn}{2470-0029},
  \urlprefix\url{http://dx.doi.org/10.1103/PhysRevD.102.043011}.

\bibitem[{\citenamefont{Martynov et~al.}(2019)\citenamefont{Martynov, Miao,
  Yang, Vivanco, Thrane, Smith, Lasky, East, Adhikari, Bauswein
  et~al.}}]{Martynov_2019}
\bibinfo{author}{\bibfnamefont{D.}~\bibnamefont{Martynov}},
  \bibinfo{author}{\bibfnamefont{H.}~\bibnamefont{Miao}},
  \bibinfo{author}{\bibfnamefont{H.}~\bibnamefont{Yang}},
  \bibinfo{author}{\bibfnamefont{F.~H.} \bibnamefont{Vivanco}},
  \bibinfo{author}{\bibfnamefont{E.}~\bibnamefont{Thrane}},
  \bibinfo{author}{\bibfnamefont{R.}~\bibnamefont{Smith}},
  \bibinfo{author}{\bibfnamefont{P.}~\bibnamefont{Lasky}},
  \bibinfo{author}{\bibfnamefont{W.~E.} \bibnamefont{East}},
  \bibinfo{author}{\bibfnamefont{R.}~\bibnamefont{Adhikari}},
  \bibinfo{author}{\bibfnamefont{A.}~\bibnamefont{Bauswein}},
  \bibnamefont{et~al.}, \bibinfo{journal}{Phys. Rev. D}
  \textbf{\bibinfo{volume}{99}}, \bibinfo{pages}{102004}
  (\bibinfo{year}{2019}),
  \urlprefix\url{https://link.aps.org/doi/10.1103/PhysRevD.99.102004}.

\bibitem[{\citenamefont{Ackley et~al.}(2020)\citenamefont{Ackley, Adya,
  Agrawal, Altin, Ashton, Bailes, Baltinas, Barbuio, Beniwal, Blair
  et~al.}}]{Ackley_2020}
\bibinfo{author}{\bibfnamefont{K.}~\bibnamefont{Ackley}},
  \bibinfo{author}{\bibfnamefont{V.~B.} \bibnamefont{Adya}},
  \bibinfo{author}{\bibfnamefont{P.}~\bibnamefont{Agrawal}},
  \bibinfo{author}{\bibfnamefont{P.}~\bibnamefont{Altin}},
  \bibinfo{author}{\bibfnamefont{G.}~\bibnamefont{Ashton}},
  \bibinfo{author}{\bibfnamefont{M.}~\bibnamefont{Bailes}},
  \bibinfo{author}{\bibfnamefont{E.}~\bibnamefont{Baltinas}},
  \bibinfo{author}{\bibfnamefont{A.}~\bibnamefont{Barbuio}},
  \bibinfo{author}{\bibfnamefont{D.}~\bibnamefont{Beniwal}},
  \bibinfo{author}{\bibfnamefont{C.}~\bibnamefont{Blair}},
  \bibnamefont{et~al.}, \bibinfo{journal}{Publications of the Astronomical
  Society of Australia} \textbf{\bibinfo{volume}{37}} (\bibinfo{year}{2020}),
  ISSN \bibinfo{issn}{1448-6083},
  \urlprefix\url{http://dx.doi.org/10.1017/pasa.2020.39}.

\bibitem[{\citenamefont{Alford et~al.}(2018)\citenamefont{Alford, Bovard,
  Hanauske, Rezzolla, and Schwenzer}}]{Alford_2018}
\bibinfo{author}{\bibfnamefont{M.~G.} \bibnamefont{Alford}},
  \bibinfo{author}{\bibfnamefont{L.}~\bibnamefont{Bovard}},
  \bibinfo{author}{\bibfnamefont{M.}~\bibnamefont{Hanauske}},
  \bibinfo{author}{\bibfnamefont{L.}~\bibnamefont{Rezzolla}}, \bibnamefont{and}
  \bibinfo{author}{\bibfnamefont{K.}~\bibnamefont{Schwenzer}},
  \bibinfo{journal}{Phys. Rev. Lett.} \textbf{\bibinfo{volume}{120}},
  \bibinfo{pages}{041101} (\bibinfo{year}{2018}),
  \urlprefix\url{https://link.aps.org/doi/10.1103/PhysRevLett.120.041101}.

\bibitem[{\citenamefont{Sekiguchi
  et~al.}(2011{\natexlab{b}})\citenamefont{Sekiguchi, Kiuchi, Kyutoku, and
  Shibata}}]{Sekiguchi_2011_2}
\bibinfo{author}{\bibfnamefont{Y.}~\bibnamefont{Sekiguchi}},
  \bibinfo{author}{\bibfnamefont{K.}~\bibnamefont{Kiuchi}},
  \bibinfo{author}{\bibfnamefont{K.}~\bibnamefont{Kyutoku}}, \bibnamefont{and}
  \bibinfo{author}{\bibfnamefont{M.}~\bibnamefont{Shibata}},
  \bibinfo{journal}{Phys. Rev. Lett.} \textbf{\bibinfo{volume}{107}},
  \bibinfo{pages}{051102} (\bibinfo{year}{2011}{\natexlab{b}}),
  \urlprefix\url{https://link.aps.org/doi/10.1103/PhysRevLett.107.051102}.

\bibitem[{\citenamefont{Perego et~al.}(2019)\citenamefont{Perego, Bernuzzi, and
  Radice}}]{Perego_2019}
\bibinfo{author}{\bibfnamefont{A.}~\bibnamefont{Perego}},
  \bibinfo{author}{\bibfnamefont{S.}~\bibnamefont{Bernuzzi}}, \bibnamefont{and}
  \bibinfo{author}{\bibfnamefont{D.}~\bibnamefont{Radice}},
  \bibinfo{journal}{The European Physical Journal A}
  \textbf{\bibinfo{volume}{55}} (\bibinfo{year}{2019}), ISSN
  \bibinfo{issn}{1434-601X},
  \urlprefix\url{http://dx.doi.org/10.1140/epja/i2019-12810-7}.

\bibitem[{\citenamefont{Rosswog and LiebendÃ¶rfer}(2003)}]{Rosswog_2003}
\bibinfo{author}{\bibfnamefont{S.}~\bibnamefont{Rosswog}} \bibnamefont{and}
  \bibinfo{author}{\bibfnamefont{M.}~\bibnamefont{LiebendÃ¶rfer}},
  \bibinfo{journal}{MNRAS} \textbf{\bibinfo{volume}{342}}, \bibinfo{pages}{673}
  (\bibinfo{year}{2003}), ISSN \bibinfo{issn}{0035-8711},
  \eprint{https://academic.oup.com/mnras/article-pdf/342/3/673/2873441/342-3-673.pdf},
  \urlprefix\url{https://doi.org/10.1046/j.1365-8711.2003.06579.x}.

\bibitem[{\citenamefont{Richers et~al.}(2015)\citenamefont{Richers, Kasen,
  O'Connor, Fern{\'{a}}ndez, and Ott}}]{Richers_2015}
\bibinfo{author}{\bibfnamefont{S.}~\bibnamefont{Richers}},
  \bibinfo{author}{\bibfnamefont{D.}~\bibnamefont{Kasen}},
  \bibinfo{author}{\bibfnamefont{E.}~\bibnamefont{O'Connor}},
  \bibinfo{author}{\bibfnamefont{R.}~\bibnamefont{Fern{\'{a}}ndez}},
  \bibnamefont{and} \bibinfo{author}{\bibfnamefont{C.~D.} \bibnamefont{Ott}},
  \bibinfo{journal}{ApJ} \textbf{\bibinfo{volume}{813}}, \bibinfo{pages}{38}
  (\bibinfo{year}{2015}),
  \urlprefix\url{https://doi.org/10.1088%2F0004-637x%2F813%2F1%2F38}.

\bibitem[{\citenamefont{Sumiyoshi et~al.}(2020)\citenamefont{Sumiyoshi,
  Fujibayashi, Sekiguchi, and Shibata}}]{sumiyoshi2020properties}
\bibinfo{author}{\bibfnamefont{K.}~\bibnamefont{Sumiyoshi}},
  \bibinfo{author}{\bibfnamefont{S.}~\bibnamefont{Fujibayashi}},
  \bibinfo{author}{\bibfnamefont{Y.}~\bibnamefont{Sekiguchi}},
  \bibnamefont{and} \bibinfo{author}{\bibfnamefont{M.}~\bibnamefont{Shibata}},
  \emph{\bibinfo{title}{Properties of neutrino transfer in a deformed remnant
  of neutron star merger}} (\bibinfo{year}{2020}), \eprint{2010.10865}.

\bibitem[{\citenamefont{Abe et~al.}(2018)\citenamefont{Abe, Bronner, Hayato,
  Ikeda, Iyogi, Kameda, Kato, Kishimoto, Marti, Miura et~al.}}]{Abe_2018}
\bibinfo{author}{\bibfnamefont{K.}~\bibnamefont{Abe}},
  \bibinfo{author}{\bibfnamefont{C.}~\bibnamefont{Bronner}},
  \bibinfo{author}{\bibfnamefont{Y.}~\bibnamefont{Hayato}},
  \bibinfo{author}{\bibfnamefont{M.}~\bibnamefont{Ikeda}},
  \bibinfo{author}{\bibfnamefont{K.}~\bibnamefont{Iyogi}},
  \bibinfo{author}{\bibfnamefont{J.}~\bibnamefont{Kameda}},
  \bibinfo{author}{\bibfnamefont{Y.}~\bibnamefont{Kato}},
  \bibinfo{author}{\bibfnamefont{Y.}~\bibnamefont{Kishimoto}},
  \bibinfo{author}{\bibfnamefont{L.}~\bibnamefont{Marti}},
  \bibinfo{author}{\bibfnamefont{M.}~\bibnamefont{Miura}},
  \bibnamefont{et~al.}, \bibinfo{journal}{ApJ} \textbf{\bibinfo{volume}{857}},
  \bibinfo{pages}{L4} (\bibinfo{year}{2018}), ISSN \bibinfo{issn}{2041-8213},
  \urlprefix\url{http://dx.doi.org/10.3847/2041-8213/aabaca}.

\bibitem[{\citenamefont{Albert et~al.}(2017)\citenamefont{Albert, André,
  Anghinolfi, Ardid, Aubert, Aublin, Avgitas, Baret, Barrios-Martí, Basa
  et~al.}}]{Albert_2017}
\bibinfo{author}{\bibfnamefont{A.}~\bibnamefont{Albert}},
  \bibinfo{author}{\bibfnamefont{M.}~\bibnamefont{André}},
  \bibinfo{author}{\bibfnamefont{M.}~\bibnamefont{Anghinolfi}},
  \bibinfo{author}{\bibfnamefont{M.}~\bibnamefont{Ardid}},
  \bibinfo{author}{\bibfnamefont{J.-J.} \bibnamefont{Aubert}},
  \bibinfo{author}{\bibfnamefont{J.}~\bibnamefont{Aublin}},
  \bibinfo{author}{\bibfnamefont{T.}~\bibnamefont{Avgitas}},
  \bibinfo{author}{\bibfnamefont{B.}~\bibnamefont{Baret}},
  \bibinfo{author}{\bibfnamefont{J.}~\bibnamefont{Barrios-Martí}},
  \bibinfo{author}{\bibfnamefont{S.}~\bibnamefont{Basa}}, \bibnamefont{et~al.},
  \bibinfo{journal}{ApJ} \textbf{\bibinfo{volume}{850}}, \bibinfo{pages}{L35}
  (\bibinfo{year}{2017}), ISSN \bibinfo{issn}{2041-8213},
  \urlprefix\url{http://dx.doi.org/10.3847/2041-8213/aa9aed}.

\bibitem[{\citenamefont{Farrow et~al.}(2019)\citenamefont{Farrow, Zhu, and
  Thrane}}]{Farrow_2019}
\bibinfo{author}{\bibfnamefont{N.}~\bibnamefont{Farrow}},
  \bibinfo{author}{\bibfnamefont{X.-J.} \bibnamefont{Zhu}}, \bibnamefont{and}
  \bibinfo{author}{\bibfnamefont{E.}~\bibnamefont{Thrane}},
  \bibinfo{journal}{ApJ} \textbf{\bibinfo{volume}{876}}, \bibinfo{pages}{18}
  (\bibinfo{year}{2019}), ISSN \bibinfo{issn}{1538-4357},
  \urlprefix\url{http://dx.doi.org/10.3847/1538-4357/ab12e3}.

\bibitem[{\citenamefont{Belczynski et~al.}(2008)\citenamefont{Belczynski,
  O{\textquotesingle}Shaughnessy, Kalogera, Rasio, Taam, and
  Bulik}}]{Belczynski_2008}
\bibinfo{author}{\bibfnamefont{K.}~\bibnamefont{Belczynski}},
  \bibinfo{author}{\bibfnamefont{R.}~\bibnamefont{O{\textquotesingle}Shaughnessy}},
  \bibinfo{author}{\bibfnamefont{V.}~\bibnamefont{Kalogera}},
  \bibinfo{author}{\bibfnamefont{F.}~\bibnamefont{Rasio}},
  \bibinfo{author}{\bibfnamefont{R.~E.} \bibnamefont{Taam}}, \bibnamefont{and}
  \bibinfo{author}{\bibfnamefont{T.}~\bibnamefont{Bulik}},
  \bibinfo{journal}{ApJ} \textbf{\bibinfo{volume}{680}}, \bibinfo{pages}{L129}
  (\bibinfo{year}{2008}), \urlprefix\url{https://doi.org/10.1086%2F589970}.

\bibitem[{\citenamefont{Lasky et~al.}(2014)\citenamefont{Lasky, Haskell, Ravi,
  Howell, and Coward}}]{Lasky_2014}
\bibinfo{author}{\bibfnamefont{P.~D.} \bibnamefont{Lasky}},
  \bibinfo{author}{\bibfnamefont{B.}~\bibnamefont{Haskell}},
  \bibinfo{author}{\bibfnamefont{V.}~\bibnamefont{Ravi}},
  \bibinfo{author}{\bibfnamefont{E.~J.} \bibnamefont{Howell}},
  \bibnamefont{and} \bibinfo{author}{\bibfnamefont{D.~M.}
  \bibnamefont{Coward}}, \bibinfo{journal}{Phys. Rev. D}
  \textbf{\bibinfo{volume}{89}}, \bibinfo{pages}{047302}
  (\bibinfo{year}{2014}),
  \urlprefix\url{https://link.aps.org/doi/10.1103/PhysRevD.89.047302}.

\bibitem[{\citenamefont{Galaudage et~al.}(2020)\citenamefont{Galaudage,
  Adamcewicz, Zhu, and Thrane}}]{heavy_dns}
\bibinfo{author}{\bibfnamefont{S.}~\bibnamefont{Galaudage}},
  \bibinfo{author}{\bibfnamefont{C.}~\bibnamefont{Adamcewicz}},
  \bibinfo{author}{\bibfnamefont{X.-J.} \bibnamefont{Zhu}}, \bibnamefont{and}
  \bibinfo{author}{\bibfnamefont{E.}~\bibnamefont{Thrane}},
  \emph{\bibinfo{title}{Heavy double neutron stars: birth, mid-life and death}}
  (\bibinfo{year}{2020}), \eprint{2011.01495}.

\end{thebibliography}

\end{document}